\documentstyle{article}
\title{
\begin{flushright}
{\bf\normalsize   COLO-HEP-340, LPTHE-Orsay-94-62}  \\
\end{flushright}
\vskip 10pt
\bf  Ising Spins on Thin Graphs
}
\author{ {\it C.F. Baillie} \\
         Dept of Computer Science, University of Colorado\\
         Boulder, CO 80309, USA\\ \\
         {\it D.A. Johnston}$^{1}$ and {\it J. P. Kownacki}\\
         LPTHE\\
         Universite Paris Sud, Batiment 211\\
         F-91405 Orsay, France
}
\date { 28 July 1994 }         
\begin{document}
  \maketitle
                      {\Large
                      \begin{abstract}
%
The Ising model on ``thin'' graphs (standard Feynman diagrams)
displays several interesting properties. For ferromagnetic
couplings there is a mean field phase transition at the
corresponding Bethe lattice transition point.
For antiferromagnetic couplings
the replica trick gives some evidence for a spin glass phase.
In this paper we investigate both the ferromagnetic
and antiferromagnetic models
with the aid of simulations. We confirm the Bethe lattice values
of the critical points for the ferromagnetic model on $\phi^3$
and $\phi^4$ graphs and
examine the putative spin glass phase in the antiferromagnetic model by
looking at the overlap between replicas in a quenched ensemble of graphs.
We also compare the Ising results with those for higher state Potts models
and Ising models on ``fat'' graphs,
such as those used in 2D gravity simulations.
\\ \\
Submitted to Nucl. Phys. B. [FS]  \\ \\
$~^{1}$ {\it Address:} Sept. 1993 - 1994, {\it Permanent Address:} Maths Dept,
Heriot-Watt
University, Edinburgh, Scotland \\
%
                        \end{abstract} }
%
  \thispagestyle{empty}
%
%
  \newpage
%
                  \pagenumbering{arabic}

\section{Introduction}
There has been a considerable amount of both analytical and numerical work on
the Ising model living on dynamically triangulated random surfaces and the
dual $\phi^3$ (and also $\phi^4$) graphs in the past few years, largely
motivated
by the interest of the model for string theory and two-dimensional gravity.
The partition function of the model for a fixed number
of points $n$ is
\begin{equation}
Z_n(\beta ) = \sum_{G^n} \sum_{\{ \sigma \}}  \ \exp \left( \beta  \
\sum_{<ij>} G^n_{ij} \sigma_i \sigma_j \right)
\label{e01}
\end{equation}
where $G^n_{ij}$ is the connectivity matrix for a given graph or triangulation
and
the sum over graphs $\sum_{G^n}$ means that we are, in effect,
looking at annealed connectivity disorder in such models.
The model was solved
in the planar limit in \cite{1} by noting that a grand canonical partition
function
could be obtained as the free energy of a
zero dimensional $N \times N$ hermitian matrix model \cite{2}
\begin{equation}
F = { 1 \over N^2} \log \int D^{N^2} \phi_{a} D^{N^2} \phi_{b}  \exp \left( -
{1 \over
2 } \phi_a K_{ab} \phi_b + V(\phi_a, \phi_b) \right)
\label{e03}
\end{equation}
in the limit $N \rightarrow \infty$. The matrix propagator in the above is
given by
\begin{equation}
\begin{array}{cc} K^{-1}_{ab} = & \left(\begin{array}{cc}
\sqrt{g} & { 1 \over \sqrt{g}} \\
{1 \over \sqrt{g}} & \sqrt{g}
\end{array} \right) \end{array}
\end{equation}
where $g = \exp (2 \beta)$,
and the potentials for $\phi^3$ and $\phi^4$ graphs are given by
\begin{equation}
V(\phi_a, \phi_b) = {\lambda \over 3 } (\phi_a^3 + \phi_b^3 )
\end{equation}
and
\begin{equation}
V(\phi_a, \phi_b) = {\lambda \over 4 } (\phi_a^4 + \phi_b^4 )
\end{equation}
respectively.
We have suppressed
the external field dependence for clarity, though
unlike the case of a fixed 2d lattice an exact
solution is still possible even in the
presence of an external field.
The diagrams generated by the matrix model integral
are ``fat'' graphs because the two indices carried by the matrix
fatten the propagators into ribbons.
This gives enough structure to build surfaces
with the trace round internal loops and the initial factors in the coupling
and propagator producing just the right factor of $N$ to count the genus.
The matrix model solution has been pushed as far as toroidal
topology \cite{3}, and both the toroidal and planar results for the
(non-Onsager)
critical exponents are in agreement with continuum derivations in
lightcone and conformal gauge
Liouville theory \cite{4}. There is now a gratifyingly large body of numerical
evidence
supporting the dynamical lattice exponents for both Ising and Potts models
\cite{5}.

Divorcing the Ising model calculations and simulations
from their origins in string theory and regarding them as an exercise
in statistical mechanics, it is natural to ask if the methods employed might
also
be applicable to {\it quenched} disorder. In such a case
the free energy $F$ is calculated
on each graph before
summing over the random graphs
\begin{equation}
F = \sum_{G} \log Z (G)
\label{e14}
\end{equation}
where $Z(G)$ is  the partition function for the Ising
model on a particular random graph. The logarithm can be dealt with
by using the replica trick \cite{6}
\begin{eqnarray}
\log Z &\simeq& {Z^n - 1 \over n} \nonumber \\
& n \rightarrow 0&
\end{eqnarray}
at the expense of having to calculate with $n$ replicas of the original matter
system. On a fixed lattice this is not a problem, but the $c=1$ barrier
in 2d gravity means that it isn't possible to calculate with more than 2 Ising
replicas. Ignoring the problem and simply taking the limit $n \rightarrow 0$
in the formulae of \cite{4} for the critical exponents, does however give
apparently reasonable results that are close to the dynamical lattice
exponents,
which is what is seen in a direct simulation of a quenched set of $\phi^3$
fat graphs \cite{7}.

It would obviously be better to have a more trustworthy formalism for
investigating
quenched problems where such troubles did not arise and one could confidently
calculate with an arbitrary number of replicas. Such a formalism is actually
inherent in the matrix model approach in a limit that is
opposite in a sense
to that usually considered, namely $N \rightarrow 1$ \cite{8}.
In this limit the fat graphs that build the surfaces or their duals
in the matrix model integrals degenerate to ``thin'' graphs - in other
words zero dimensional Feynman diagrams.
Such graphs have in fact been considered before in the context of
spin glass theories \cite{8a}, without utilizing the methods of field
theory and large order expansions of \cite{8}.
The generic problems in the matrix model
approach for $c>1$ can be traced to the difficulty in evaluating angular
integrals that arise from diagonalizing the matrices, and these disappear for
$N=1$. Consequently the difficulties in applying the replica formalism also
vanish.
As the regular surface interpretation is lost \footnote{One can,
however, think of the
thin graphs as ``infinite genus'' surfaces with a maximal handle
density \cite{8}.} the continuum limit is taken by forcing the
number of points $n \rightarrow \infty$ rather than tuning the
cosmological constant as for $N \rightarrow \infty$.

In the remainder of the paper
we discuss the behaviour of the Ising model on such thin graphs.
In the next section we review the analytical calculations of \cite{8}
for ferromagnetic couplings, which serve as a test
of the method.
In this context
we also discuss the saddle point solution for higher $\phi^n$
interactions and $q=3,4$ state Potts models and
their close correspondance with Bethe lattice results.
We then
describe in section 3 simulations of the ferromagnetic Ising model
on $\phi^3$ and $\phi^4$ graphs,
carried out essentially as a test
of the formalism
and the simulation programs, which are in agreement with these
results. In section 4 we discuss the {\it anti}-ferromagnetic
Ising model on thin graphs and describe our simulations of
both the Ising and three and four state Potts models with
purely negative couplings on quenched ensembles of Feynman diagrams.
The calculations of \cite{8} suggest that
the antiferromagnetic model may have a spin glass phase at sufficiently
negative coupling, and we investigate this numerically by histogramming
the overlaps between replicas.
In sections 5 and 6 we discuss briefly some exploratory
simulations of the Ising antiferromagnet on quenched fat graphs
and of Potts antiferromagnets on thin graphs respectively, largely
to compare the results with those for the Ising model.
Finally in section 7
we present our conclusions and discuss the various loose ends
that have been left untied by the current work.

\section{Analytical Calculations on Thin Graphs}
The (annealed) average of the
partition function over all the thin graphs with $2n$ vertices can
be expressed as
\begin{equation}
Z_n(\beta) \times N_n = {1 \over 2 \pi i} \oint { d \lambda \over
\lambda^{2n + 1}} \int {d \phi_a d \phi_b \over 2 \pi \sqrt{\det K}}
\exp (- S )
\end{equation}
where $N_n$ is the number of undecorated graphs with $2n$ points,
\begin{equation}
N_n = \left( {1 \over 6} \right)^{2n} { ( 6 n - 1 ) !! \over ( 2 n ) !!
}
\end{equation}
for $\phi^3$ graphs, and the action is identical to that of
the matrix model in equ.(2).

If we rescale the fields $\phi \rightarrow \phi / \lambda$ and drop
the overall factors that are not germane to the discussion then
the large $n$ limit can be evaluated by calculating the saddle
points of the action
\begin{equation}
S = {1 \over 2 } ( \phi_a^2 + \phi_b^2) - { 1 \over g} \phi_a \phi_b -
{1 \over 3} (\phi_a^3 + \phi_b^3).
\end{equation}
A non-trivial phase structure can appear
if there is a crossover in the dominant saddlepoint values,
and this is what does in fact occur. In the $\phi^3$
model the solution to the saddle
point equations are
\begin{eqnarray}
\phi_a,\phi_b &=&  { ( g - 1 )  \over
         g}  \nonumber \\
\phi_a,\phi_b &=& { 1 + g \pm \sqrt{ (g + 1) ( g  - 3 ) }
\over 2   g  }
\end{eqnarray}
(along with a trivial zero solution, and a solution exchanging
$\phi_a,\phi_b$ in the second case). The first set
of solutions above is the high temperature unmagnetized phase
and the second, which becomes dominant for $g>3$ is the low
temperature magnetized phase, with the magnetization
being given by \cite{8}
\begin{equation}
M = { \phi_a^3 - \phi_b^3 \over \phi_a^3 + \phi_b^3}.
\end{equation}
If we denote the deviation from the critical coupling $\beta_c$ by
$t = | \beta - \beta_c | / \beta_c$
then we see that we have mean-field like behaviour $M \simeq
t^{1/2}$. The same is true for the other exponents which
can easily be calculated using the results of \cite{8}.

The saddle point equations can also be readily solved for the
case of the Ising model on $\phi^4$ graphs, to give
\begin{eqnarray}
\phi_a,\phi_b &=& - \sqrt{ 1 - 1 / g } \nonumber\\
\phi_a &=& { \left( \sqrt{ 1 - \sqrt{1 - 4 / g^2}}\right)
\left( 1 + \sqrt{1 - 4 / g^2} \right) g \over  2 \sqrt{2}},
\nonumber\\
\phi_b &=& { \sqrt{ 1 - \sqrt{1 - 4 / g^2}} \over \sqrt{2}}.
\end{eqnarray}
which indicate a phase transition at $g=2$, again with mean field
exponents. If we pause for a second, and look back over the results
we can see that we have reproduced the behaviour of the Ising model
on Bethe lattices
with 3 and 4 neighbours \cite{9} with these results including, remarkably,
the relevant critical temperatures. The correspondence between the
Bethe lattice results and those on Feynman diagrams is unexpected
because the Bethe lattice contains no loops, whereas the number of loops
in the Feynman diagrams which can be calculated using standard methods
is extensive \cite{10}, being $n + 1$ for $\phi^3$ graphs with $2n$ points for
instance. The correspondance is not a fluke, it persists for
the Ising model on $\phi^{5,6..}$ graphs and the corresponding Bethe
lattices. Indeed, it remains true for the 3 and 4 state Potts models
also, where we have
\begin{equation}
g_c = { z + p -2 \over z -2}
\end{equation}
for the critical value of $g$
in a $p$-state Potts model on a Bethe lattice with $z$ neighbours \cite{11}.
The action for the 3 state Potts model on
$\phi^3$ graphs
is just
\begin{equation}
S = { 1 \over 2} ( \phi_a^2 + \phi_b^2 + \phi_c^2 ) - c (\phi_a \phi_b + \phi_b
\phi_c
+ \phi_c \phi_a) - { 1 \over 3} (\phi_a^3 + \phi_b^3 + \phi_c^3)
\end{equation}
where $c$ is now
given by
\begin{equation}
c = { 1 \over g + p -2}
\end{equation}
with $p=3$.
This gives the following
high and low temperature saddle point solutions
\begin{eqnarray}
\phi_a , \phi_b , \phi_c &=& 1 - 2 c \nonumber \\
\phi_a &=& { 1 + c + \sqrt{1 - 4 c - 3c^2 - 2 c \sqrt{ 1 - 4 c - 4c^2}} \over
2}
\nonumber \\
\phi_b &=& { 1 + c - \sqrt{1 - 4 c - 3c^2 - 2 c \sqrt{ 1 - 4 c - 4c^2}} \over
2}
\nonumber \\
\phi_c &=& { 1 + \sqrt{ 1 - 4 c - 4c^2} \over 2}
\end{eqnarray}
(along with permutations) which again display a transition at the Bethe lattice
value of $g_c=4$ (ie $c = 1 / 5$). At the risk of belabouring the point the
high and low temperature
solutions for the 4 state Potts model on $\phi^3$ graphs
\begin{eqnarray}
\phi_a , \phi_b , \phi_c , \phi_d &=& 1 - 3 c \nonumber \\
\phi_a , \phi_b , \phi_c , \phi_d &=& { 1 + c \pm \sqrt{ 1 - 6 c - 7c^2} \over
2},
\end{eqnarray}
where two of the signs are to be taken positive and two negative
in the second low temperature set,
also fall into the pattern with $g_c =5$ ($c = 1 / 7$).

It is thus clear that, at least for ferromagnetic bonds where there
is no frustration, the Feynman diagrams offer an effective
way of calculating with a closed version of the Bethe lattice.
In fact, they might even be considered as more respectable than
the Bethe lattice itself where one has to throw away the terms
in the outermost shells to get a phase transition \cite{9}
\footnote{The lattice {\it with} the boundary terms included
is often called the Cayley tree, and its thermodynamic behaviour is
dominated by the terms on the boundary.}. The correspondance between
the diagrams and the Bethe lattice is underlined by the following
observation, due to H. Hilhorst \cite{9a}: We can construct one of our
Feynman diagrams in a Bethe-lattice-like manner by
picking a seed point and following the
standard procedure for constructing a Bethe lattice, but also
allowing ourselves the possibility of joining branches to other
branches
at any step. This then gives the correct loop counting of $n + 1$ in
the $\phi^3$ model with $2n$ vertices, once all the branches have been closed.

\section{Ferromagnet: Methodology and Simulations}

As a warmup exercise for simulations of the antiferromagnetic
Ising model on Feynman diagrams we did some smaller scale simulations
of the ferromagnetic model on $\phi^3$ and $\phi^4$ graphs.
The generation of the graphs is a much easier exercise than generating
the planar $\phi^3$ graphs used in $2d$ gravity simulations as the
absence of a constraint on the topology obviates the need to perform
gymnastics with Tutte's algorithm. The simulation itself used standard
finite size scaling techniques to determine the critical exponents.
The absence of frustration in the ferromagnetic model allowed the use
of the Wolff algorithm to give effective spin updating. We simulated on
$\phi^3$ and $\phi^4$ graphs
of sizes $250,500,1000,2500$ and $5000$
points for various $\beta$ values, performing
20,000 measurements (after a suitable number of metropolis
equilibration sweeps) with
the number of Wolff sweeps per measurement
being determined by the inverse of the mean cluster size.
With ferromagnetic couplings and no frustration the quantities
we use to characterize the transition, such as the energy and
magnetization are all self-averaging, so we might expect to be able
to obtain the correct critical behaviour using a {\it single} graph of
each size, provided our graphs are not too small and we pick
a representative graph. We shall see that this is, in fact, the case.

We commence our analysis of the results by looking at the crossing
of the (unnormalized) Binder's cumulant for the magnetization
\begin{equation}
U_M =  { <M^4> \over <M^2>^2}
\end{equation}
to locate the critical point. This provides a good test that we
have managed to generate our graphs correctly.
If we consider $\phi^3$ graphs first, the critical
value of $g=3$ translates to $\beta_c \simeq 0.549$ and the crossing of the
cumulant gives the estimate $\beta_c \simeq 0.547(2)$.
Similarly on $\phi^4$ graphs
$g=2$ translates to $\beta_c \simeq 0.347$ and the crossing of the
cumulant gives the estimate $\beta_c \simeq 0.345(4)$.
The quoted errors are the differences in the crossing points
for different ratios of graph sizes. As a check we have done
two complete sets of simulations with a different random graph
of each size, and used all the possible combinations
in the cumulant analysis. There was no appreciable variation
between different graphs of the same size. The crossing of the
cumulant is shown in Fig3.1 for three of the graph sizes used.

The value of the specific heat critical
exponent $\alpha$ for the mean field Ising model is
zero, signifying a discontinuity between a finite value
at $\beta_c+$ and zero at $\beta_c-$ rather than the
logarithmic divergence of the Onsager solution. If we look
at Fig3.2 we see clear signs of this mean field behaviour
in the form of the specific heat curve
and a direct fit to the data is also consistent with $\alpha=0$.
We can verify this by looking at the scaling of the
maximum slope of the magnetization cumulant which is
expected to behave as
\begin{equation}
max \left( { d U_{M} \over d \beta } \right) \simeq n^{1 \over \nu d}
\end{equation}
for a graph with $n$ vertices.
This gives $\nu d = 2.00(5)$ which, using the hyperscaling relation
$\alpha = 2 - \nu d $, is again consistent with $\alpha=0$.

The value of $\nu d$ can also be fed into the finite
size scaling relation for the vanishing of the
magnetization at the critical point
\begin{equation}
M \simeq n^{-\beta / \nu d}
\end{equation}
to extract a value for $\beta$. We find that $\beta / \nu d = 0.24(1)$,
so $\beta \simeq 1 / 2$
which is the mean field value. Similarly,
the scaling of the linear susceptibility $\chi_M$ at the critical point
\begin{equation}
\chi_M = { d M \over d H} \simeq n^{\gamma / \nu d}
\label{cm}
\end{equation}
can be fitted to extract the exponent $\gamma$, giving $\gamma = 1.06(3)$
which is in respectable agreement with the mean field value of 1.

We spare the reader the extraction of further exponents and the analysis
of the $\phi^4$ results in a similar vein, as it is clear from the
results already presented that the Ising model exhibits a mean field
transition (as expected). The cumulant analysis also confirms that
the transition occurs at the corresponding Bethe lattice transition
points. Fortified by the good agreement between the analytical
calculations and our simulations, we can now proceed to the case that
really interests us, the antiferromagnet, to look for spin glass
behaviour.

\section{Antiferromagnet: Calculations and Simulations}
In the absence of a $c=1$ barrier, we are at liberty to calculate with
as many replicas as we can manage on thin graphs. It was pointed out
in \cite{8} that the saddle point equations for $k$ replicas of the
Ising model on $\phi^3$ graphs admitted a factorized solution of the
form
\begin{equation}
\Phi = 2^{1 -k} \phi_1 \ \phi_2 \ ... \ \phi_k
\end{equation}
where the $\phi_i$ are individual solutions of the annealed, $k=1$,
equation. The $k \rightarrow 0$ quenched free energy was
equal to the annealed free energy, modulo finite size corrections.
However, the entropy per spin calculated with this solution
became negative for $g \simeq 0.211$ and higher moments of the
partition function showed that the factorized solution no longer
dominated
\begin{equation}
\lim_{n \rightarrow \infty} { 1 \over 2 n }  \log \overline { Z^k }
\ne
\lim_{n \rightarrow \infty} { k \over 2 n }  \log \overline Z
\end{equation}
with the changeover point apparently converging to a
finite value for $k$ increasing. Spin glass transitions
are often signalled by the breakdown between quenched and annealed
averages so, while this is not conclusive evidence for such a transition,
it is certainly suggestive.

Despite the fact that only antiferromagnetic couplings are present in the
model,
there is certainly frustration due to the loops. Indeed, a variational
calculation by de Seze some years ago \cite{12} noted that an
Ising antiferromagnet with dilute bond disorder {\it would}
have a spin glass phase if the undiluted version of the model
did not allow antiferromagnetic order at $T=0$. The hamiltonian in
this case was
\begin{equation}
H = - \sum_{ij} J_{ij} \sigma_i \sigma_j
\end{equation}
where the sum was over all $i,j$ and the $J_{ij}$ were
distributed (in a quenched ensemble) according to
\begin{equation}
P(J_{ij}) = c \ \delta ( J_{ij} + J) + ( 1 - c) \ \delta ( J_{ij})
\end{equation}
which has the effect of killing a fraction $1-c$ of the bonds
in any given realization. The model we consider on Feynman
diagrams could be cast in rather similar form, as we have
some probability that two vertices are connected for a given graph,
so viewed in this light spin-glass like behaviour is not so
surprising.

Although we would no longer expect such close similarity with the
standard Bethe lattice as frustration has highlighted the role of
the loops, there have been previous simulations of ``closed''
versions of the Bethe lattice \cite{13}
that might be expected to be a closer
approximation. These considered mixed ferromagnetic and
antiferromagnetic bonds on a Bethe lattice where the branches
were closed in two ways - either with fixed boundary conditions,
where no spin glass behaviour was seen, or with pairing
between different branches in a given shell,
which did display signs of a spin glass phase.
Unfortunately purely antiferromagnetic couplings
were not considered, which would have allowed a more
direct comparison with our results as
the Feynman diagrams are effectively a multi-shell generalization of
the paired closure.

Our tactics for simulating the antiferromagnet were rather
similar to those in \cite{13}.
We took an explicitly quenched average by generating
$O(100)$ different Feynman diagrams for each of the sizes
that we simulated, $100,250,500$ and $1000$ points.
The number of lattices simulated is justified {\it a posteriori}
by checking that the results for non-self averaging observables,
such as the overlap described in the next paragraph,
look reasonable.
We used
a massively parallel processor (the Intel Paragon)
to run the different lattices
concurrently and perform the requisite
quenched averaging over the observables during
the course of the simulation. For the sake of simplicity
we made no attempt to use either multi-canonical,
simulated tempering, or replica cluster simulation schemes
which are applicable,
usually with some effort,
to frustrated systems such as ours \cite{14}
but stuck to the metropolis algorithm. At each $\beta$
point we employed
500,000 simulated annealing sweeps to cool the system
followed by 200,000 production sweeps with a measurement
every 10th sweep. Each sweep consisted of a complete metropolis
update of the lattice.

In addition to all of the
standard thermodynamic variables that we measured in the ferromagnetic
case
we paid particular attention to the overlap between two replicas
of the Ising model on a given graph, as non trivial behaviour
in the overlap distribution is often the sign of a spin glass phase.
The overlap is defined in the standard fashion by putting two
independent replicas on {\it each} graph in the ensemble
\begin{equation}
q = { 1 \over n} \sum_{i=1}^n \sigma_i \tau_i
\end{equation}
where $\sigma_i$ is the spin at site $i$ in one replica, and
$\tau_i$ is the spin at site $i$ in the other.
We then measure
\begin{equation}
P_n(q) = \left[ \langle \delta ( q - 1/ n \sum \sigma_i \tau_i) \rangle \right]
\end{equation}
where $[ \ \ ]$ denotes the quenched graph average and $< \ \ >$ the thermal
average, to get a histogram of the overlap probability.

$P(q)$ is
known to be strongly non-self-averaging in soluble models
such as the Sherrington Kirkpatrick (SK) model \cite{14a} so the quenched
average is vital, given the  strong variation between individual
graphs. A useful heursitic check in the simulations in to make sure
that $P(q)$ is symmetric about $q=0$ as $P(q) = P(-q)$ analytically. If this is
not
the case the data is almost certainly not equilibrated. We have also measured
the autocorrelation times in the measurements of $P(q)$ in order to help judge
their reliability.

It is known in the SK model
that $P(q)$ is related to the Parisi order parameter
by $P(q) = dx (q) / d q$ \cite{15}, but in
principle there are an infinite number of
order parameters generalizing the overlap
in short range models \cite{16}. It has also been observed
that $P(q)$ can produce misleading results in short range
spin glasses, which might not exhibit the coexistence of
infinitely many pure states that is seen in the mean field
solution of the SK model \cite{17}, but rather have a two
phase ``droplet'' structure. Recent
numerical work \cite{17a} on the nearest neighbour
four dimensional $\pm J$ Ising spin glass, however, seems
to support a mean field like picture
so the question is far from settled. In any case, given that the genuine
spin glass with nearest neighbour
$\pm J$ couplings on the Bethe lattice behaves
in an SK like way with the appropriate boundary conditions \cite{18},
it is not unreasonable to expect that $P(q)$ may be an appropriate
tool to use for the Feynman diagrams we consider here.
In addition, observables such as the spin glass susceptibility
\begin{equation}
\chi_{sg} = { 1 \over n} \sum_{ij} [ \langle \sigma_i \sigma_j
\rangle^2 ]
\label{csg1}
\end{equation}
may be calculated from the moments of $P(q)$
\begin{equation}
\chi_{sg} = n \int q^2 P_n ( q ) dq.
\label{csg2}
\end{equation}
We also measured directly the Binder's cumulant associated with $q$
\begin{equation}
U_{sg} =  { \langle q^4 \rangle \over \langle q^2
\rangle^2 }
\end{equation}
whose crossing points for differently sized lattices might be expected
to pinpoint any spin glass transition point, in a similar manner to the
cumulant for the magnetization in ferromagnetic transitions.

As in the ferromagnetic case, we commence our analysis by looking
for the crossing point of the appropriate cumulant, $U_{sg}$.
The results for this are much less accurate than for the ferromagnetic
transition, both because of the smaller lattice sizes simulated
and the non-self averaging nature of quantities that concern the
overlap. Nonetheless, we obtain a crossing for three
of the four lattice sizes, indicating
that there is a spin glass transition with
$\beta_c \simeq -0.94(2)$, where the error is again from the
different intercepts for different ratios of lattice sizes.
This translates to $g \simeq 0.15$, which is rather smaller
than the value indicated by the negative entropy
in \cite{8} ($g \simeq 0.211$ or $\beta_c \simeq -0.78$).
However, using the high temperature solution to extract the
entropy in the SK model \cite{17} gives an expression
that becomes negative {\it inside} the spin glass phase rather
than at the transition point as in the random energy model
\cite{17b}. In \cite{8} the entropy
changes sign
in what seems to be the paramagnetic phase from our simulations,
but the apparent inversion here with respect to the SK model may
be due to the purely antiferromagnetic couplings.
We show the crossing of the cumulant in Fig4.1, where we have
dropped the results from the $N=500$ lattices that (narrowly)
failed to cross the others in the plot.

It is known that the specific heat displays a cusp at
$\beta_c$ in the SK model
corresponding to an exponent $\alpha = -1$, but if we look at our
data in Fig4.2 we see that, although there is a cusp it
appears to be in the region of $\beta \simeq -0.84$.
However, the specific heat estimations of the critical coupling
need to be extrapolated to infinite volume to
compare with the cumulant results. To do this we need to
extract a value for the combination $\nu d$ from
our data \footnote{Although $d$ may be rather ill-defined
on our diagrams, the combination is not.}
which can be done by regarding the scaling of the maximum slope
of $U_{sg}$ which, just as for $U_{M}$, is expected to behave as
\begin{equation}
max \left( { d U_{sg} \over d \beta } \right) \simeq n^{1 \over \nu d}.
\end{equation}
We find $\nu d = 2.8(2)$, which is consistent with the cusp
displayed by the specific heat as the hyperscaling relation
$\alpha = 2 - \nu d$ still holds. It is also
compatible with the mean field SK value $\alpha = -1$.
If we now use our value of $\nu d$ to extrapolate the $\beta_c$'s
for the specific heat to infinite volume we find $\beta_c(\infty) \simeq
-0.88(2) $,
which is less negative than the value from the cumulant crossing. The quoted
error for the critical coupling determined from the specific
heat cusp is probably too optimistic, given it is the result of a fit
first to $\nu d$ followed by a linear interpolation
of the estimated cusp positions, so the
cumulant crossing value is certainly more reliable.

The linear susceptibility $\chi_M = dM / dH$ does not diverge
at a spin glass transition point, but is expected to display a cusp
similar to that in the specific heat. In Fig4.3 we plot
$\chi_M$ and see clear evidence of a cusp although it is noteworthy that
this is considerably displaced from $\beta \simeq - 0.95$. The finite
size scaling does not give such a convincing shift back towards this value
as in the specific heat case, but the quality of the data is poorer.
{}From the
spin glass point of view the more interesting susceptibility
is $\chi_{sg}$ defined in equ.(\ref{csg1}), which we measured
both directly and using using equ.(\ref{csg2}).
The data from the direct measurements is plotted
in Fig4.4.
$\chi_{sg}$ is expected to diverge
at a spin glass transition, in contrast to $\chi_M$, so there
will be a finite size scaling relation of the form
\begin{equation}
\chi_{sg} \simeq n^{\gamma / \nu d},
\end{equation}
analogously to equ.(\ref{cm}).
Fitting the data from both ways of extracting $\chi_{sg}$
we find $\gamma  = 1.3(2)$, where the
error bar does not take account of the uncertainty in determining
the critical point. This value is still marginally
compatible with the SK value of 1.

Looking directly at the histograms of $P(q)$ also gives
indications of a transition around
$\beta \simeq -0.95$.
In Figs4.5-4.6, we show the histograms for
$\beta=-0.1$,
deep in the paramagnetic phase
and at $\beta=-1.2$, below the transition point.
We have plotted the histograms without the error bars coming from the quenched
average (which are not particularly large) for clarity.
The histograms in Fig4.5 are clearly gaussian in shape, as expected for
a paramagnetic phase and
the width of the gaussians for differently size graphs is in
reasonable agreement with ``delta-function'' like behaviour,
namely $\sigma^2 \simeq n^{-1}$.
As the region around $\beta \simeq - 0.95$ is approached
the histograms broaden considerably and begin to
show signs of a dip at the origin.
Once we are in the putative spin glass phase the
histograms look generically like that in Fig4.6
for $\beta=-1.2$. Comparing with histograms
for the different lattice sizes
shows that the peak at $q \simeq 0.5$
is sharpening with increasing lattice size and that the
``tail'' stretching down to $q=0$ remains stable.
These results suggest that the behaviour is strikingly
similar to the SK spinglass, with a delta function at finite
$q$ adjoined to a continuous tail down to $q=0$ in
the infinite volume limit.
This also means that the distribution of $P(q)$ takes
on a form very similar to that seen on a Bethe lattice spin-glass
with closed boundary conditions in the simulations in \cite{13}.
Once we are some distance below $\beta \simeq - 0.95$
the autocorrelation times for the overlap become very large
and the histograms begin to lose their symmetry and
roughen, suggesting that
the data has failed to equilibrate, a behaviour often seen
in spin glass simulations.

{}From these results
the Ising antiferromagnet
on Feynman diagrams would appear
to possess a spin glass phase at sufficiently negative $\beta$
with similar properties to the SK model.
It is thus behaving in a mean field like manner and all the
paraphernelia of replica symmetry breaking might be expected to be applicable.

\section{Antiferromagnet on Fat Graphs}
As we indicated in the introduction there is nothing to stop us simulating
the Ising model on a quenched ensemble of fat graphs, even though such a system
is currently not accesssible to reliable analytical calculations.
The simulations
of an Ising ferromagnet on such an ensemble in \cite{7} encountered no
obvious pitfalls and gave results that were very similar to the
annealed ensemble of dynamical graphs.
One might expect quenched and annealed averages
to give rather similar results when frustration is not playing a major role, so
the antiferromagnet may well be a different story entirely.
A simulation of an Ising antiferromagnet on quenched fat graphs is of interest
because the graphs involved are no longer of an ``infinite-dimensional'' mean
field type.
Indeed, it appears that the fractal dimension
of the graphs is 3 or 4 (depending on the method of definition) \cite{19b},
and 3 may be the lower critical
dimension of the spin-glass transition. It might be possible to
look analytically at the change, if any, from the mean field behaviour on $N=1$
thin graphs as the size of the matrices $N$ is increased to fatten the graphs.
Such an approach has already provided surprisingly good values
for critical exponents and couplings in the one matrix model by extrapolation
from quite small $N$ \cite{19c}.

We are less ambitious here and simply conduct a trial simulation of
the Ising model on $O(100)$ planar $\phi^3$ graphs with 250 vertices in each.
The loop distribution is calculable using the methods of \cite{2}
and is strongly peaked at small (3,4..) loops.
This means that by comparison with
the thin graphs, where the loops are in general
rather large and the local structure
tree-like, frustration is likely to play a stronger role.
We use the same number of graphs and simulated annealing sweeps
in the simulation as in the thin graph simulations of the preceding
section.
We find that the histograms of $P(q)$ begin to deviate strongly
from a gaussian shape in the region of $\beta = -1.5$ and become
very irregular and assymetric for
$\beta < -2.0$.
This is a signal either that we have not equilibrated the system
or failed to include enough graphs in the quenched average
for these $\beta$ values. Increasing
the simulated annealing steps by a factor of four makes little
difference to the roughness of the $P(q)$.
The associated autocorrelation times for the measurements
of $q$ also rapidly become astronomical below $\beta = -1.5$.

In the region of $\beta = -1.5$ the observed
histograms are rather similar to those in
the region of $\beta = -0.95$ on thin graphs,
being perhaps more irregular.
This would mean that one has to work
harder on fat graphs than on thin graphs to obtain
reliable numerical results.
One of the  histograms for $P(q)$ at $\beta=-1.9$
for the fat graph is shown in Fig5.1, where we have
included a histogram of $P(q)$ at $\beta=-1.3$
on a thin graph of 250 vertices for comparison. Allowing for
the differing ``distances'' from the apparent critical values
and remembering that a size of 250 on a fat graph may not be equivalent
to a size of 250 on a thin graph, the two distributions are very similar
in form.

As we have only simulated a single graph size,
we make no attempt to conduct a
serious scaling analysis with our current fat graph data
and content ourselves with some qualitative observations:
The non-trivial distribution of $P(q)$ that has been observed,
and the large autocorrelation times
suggest that something non-trivial is still going on for
the antiferromagnet on fat graphs for sufficiently
negative $\beta$. If the form of $P(q)$ persisted
on larger graphs it would be suggestive of mean field like
behaviour, given its similarity to the thin graph distribution and
hence that of the SK model.

\section{Higher State Potts Models}

We have thus seen that the Ising antiferromagnet on $\phi^3$ graphs
displays a non-gaussian $P(q)$ for sufficiently negative couplings,
which strongly resembles those seen in \cite{13} for the Ising spin
glass on a Bethe lattice with closed boundary conditions, and those
for the SK model. It would appear that the model is closer
to these mean-field systems than short range spin glasses.
Given this it is of interest to enquire whether higher state
Potts models also display non-trivial $P(q)$ on Feynman
diagrams for purely negative couplings. It is not immediately
obvious that they will, as it is easier to ``unfrustrate'' a given loop
if we have more colours. In addition, genuine Potts glasses
($\pm J$ or Gaussian distributed couplings) have a
completely different order parameter distribution
to the Ising spin glass in both infinite range
models and on Bethe lattices
\cite{20,20a}. In the infinite range (mean field)
case they display two transition temperatures, with a
$P(q)$ that is still a delta function below the first and two
delta functions joined by a continuous portion below the second.
On the Bethe lattice the form of $P(q)$ depends on the bond
distribution and the number of neighbours, for $\pm J$ bonds
in the 3 state Potts model
for instance there is a delta-function when
the number of neighbours $z \ge 4$
and a continuous distribution for $z=3$.
For the 4 state Potts model there is only
a delta function \cite{20a}.
For more than four states in both cases the phase transition displays
some first order properties.

For p-state Potts models the overlap
can be defined as
\begin{equation}
q = { 1 \over n} \sum_{i=1}^n ( p  \delta_{\sigma_i , \tau_i} - 1 )
\end{equation}
which is arranged to be zero
in the uncorrelated case $\delta_{\sigma_i , \tau_i} \simeq
1 / p$. This can be measured just as in the Ising simulations.
A quick test simulation of the 3 state model
on $\phi^3$ graphs reveals that the
distribution of $P(q)$ does not appear to change from the paramagnetic form
even for large negative
couplings, suggesting there is no spinglass phase at all.
The intuition that there is now enough liberty to relieve the frustration can,
be substantiated by Brook's theorem from graph theory \cite{21}:
If G is a connected graph with maximal vertex valency $z>2$,
and if G is not complete,
then the chromatic number $\chi$
\footnote{Not to be confused with either
of the susceptibilities $\chi_M, \chi_{sg}$ earlier!},
the minimum number of colours it takes so that
no adjacent vertices are the same colour, of G is $z$
\footnote{We would like to thank N. Gilbert for pointing this out to us.}.
In this language we have a 3-regular graph that is not complete,
so $\chi=3$
and we can avoid frustration completely with 3 colours.
By the same token, we would not expect any spinglass like behaviour on $\phi^z$
graphs
for $p \ge z$ state Potts models with purely negative couplings
as $\chi=z$ for these.

Chromatic numbers for more general random graphs are rather large,
though some asymptotic bounds are known, that are
in accordance with simulations of antiferromagnetic
Potts models \cite{21a}.
There are of course, particular cases such as
the solution of the celebrated four colour problem,
which states that $\chi=4$ for any planar map \cite{22}.
$\chi=2$ trivially for a planar square
lattice and a three
colouring algorithm also exists as a consequence of the solution of
the six vertex model \cite{22a}.
Surprisingly, another
quick test simulation for a 3 state Potts model
with purely negative couplings on $\phi^4$ Feynman diagrams
(so neither Brooks' theorem
nor the four-colour theorem apply) also produced
distributions that did not
change from the paramagnetic form
with decreasing beta.
It
would be interesting to confirm this behaviour and explore it further
for different $p$ and graphs.

A brief summary of whether or not a spin glass transition appears to
occur in the cases
discussed above ($\surd$) or not ($\times$) is given in the table below.
We have left the $p=3$ model on thin $\phi^4$ graphs undecided.

\begin{center}
\begin{tabular} {|c|c|c|c|}
\hline
$p=2$ & thin & $\phi^3$ & $\surd$ \\
\hline
$p=2$ & fat & $\phi^3$ & $\surd$  \\
\hline
$p=3$ & fat,thin&  $\phi^3$ & $\times$ \\
\hline
$p=3$ & thin & $\phi^4$ & ? \\
\hline
$p=4$ & fat &  any & $\times$ \\
\hline
$p \ge z$ & fat,thin & $\phi^z$ & $\times$ \\
\hline
\end{tabular}
\end{center}
\vspace{.1in}
\centerline{{\bf Table 1}: Summary of whether a spin glass transition
appears $\surd$ or not $\times$.}
\bigskip
\section{Conclusions}

We have seen analytically in section 2 that both Ising and higher state
Potts models on Feynman diagrams reproduce the Bethe lattice
mean field behaviour, even down to the same critical couplings
for Bethe lattices with the same number of nearest neighbours.
In section 3 we saw that
the self averaging nature of the quantities that characterize the
transition, such as
the energy and the magnetization,
meant that we could obtain numerically respectable results
with a single graph. These were in good accord with the
analytical calculations. The net conclusion is thus that,
at least as far as ferromagnetic couplings are concerned,
the Ising and Potts models do not ``see'' the loops that
make the difference between the Bethe lattice and the
Feynman diagrams.

These loops {\it do}, however, play a role once antiferromagnetic
couplings are introduced. The calculations of \cite{8},
suggesting a spinglass phase
in the Ising model for sufficiently negative coupling,
are supported by the non-trivial $P(q)$'s that were observed
in section 4. The quenched average over different Feynman
diagrams was obligatory in this case because of the non-self-averaging
properties of $P(q)$. The results seen on the Feynman diagrams
are strongly reminiscent of the Bethe lattice with closed
boundary conditions that was simulated in \cite{13}.
The form of the $P(q)$ for the Ising antiferromagnet
suggests that we are closer to a mean field picture
of spin glass behaviour than a droplet one on Feynman diagrams,
again much like the Bethe lattice, and in accordance with
earlier calculations for finite connectivity random lattices.
The critical exponents we extracted were also all in reasonable
agreement with those in the mean field SK model.

In section 5 we looked briefly at the
Ising antiferromagnet on planar or fat graphs,
finding that the distribution of $P(q)$ was rather
similar to that on a thin graph of a similar size.
The antiferromagnetic
higher state Potts models that we also briefly looked at
in section 6
show interesting behaviour, and there are some
intriguing links with colouring problems in graph theory.

The utility of Feynman diagrams  as a tool in analytical
calculations with random graphs of finite connectivity
is thus clear. The close connection between the Feynman diagrams
and the Bethe lattice is somewhat unexpected at first sight, but persists
even for the antiferromagnet and closed versions of the Bethe lattice.
Both the ferromagnet and antiferromagnet have displayed essentially
mean field behaviour on thin graphs. The simulations of \cite{7}
suggest that this is {\it not} the case for the ferromagnet on fat
graphs, so it would be worth pursuing the simulations of the
antiferromagnet on fat graphs to look for non mean field behaviour
and exponents.
We have not explored issues such as the scaling of $P(q)$, the
ultrametricity (or otherwise) of the distribution of states
and the response to external fields
for the antiferromagnet in the current work, but we hope
to return to these in future.

There are also various direct generalizations
that we have not had the time to explore in the current paper.
A first is to
look at ``real'' spinglasses with either $\pm J$ or gaussian couplings
on the Feynman diagrams to compare with previous Bethe lattice results
for such models. A second obvious extension is to look at other
models such as the $XY$ model on thin graphs, for which some
analytical results already exist in the unfrustrated case \cite{23}.
It might also be worthwhile to implement some of the more
advanced simulation algorithms on the models described here.
There are clearly plenty of interesting questions to be addressed
regarding quenched disorder on both thin and fat graphs.

\section{Acknowledgements}
The ferromagnetic
simulations reported in this paper were carried out on workstations
at Heriot-Watt University, LPTHE Orsay and the University of Colorado.
The antiferromagnetic simulations were carried out
on the Front Range Consortium's
208-node Intel Paragon located at NOAA/FSL in Boulder.
Some of the Binder's cumulant analysis was carried out using
programs written by A. Krzywicki.
DAJ was supported at Orsay by an EEC Human Capital and Mobility
Fellowship and an Alliance grant and would like to thank A. Krzywicki
and H. Hilhorst for various useful conversations there.
CFB is supported by DOE under
contract DE-FG02-91ER40672, by NSF Grand Challenge Applications
Group Grant ASC-9217394 and by NASA HPCC Group Grant NAG5-2218.

\vfill
\eject
\centerline{\bf Figure Captions}
\begin{description}
\item[Fig. 3.1] The crossover in Binder's cumulant
for the magnetization in the ferromagnetic model
for three representative $\phi^3$ graphs.
\item[Fig. 3.2] The specific heat in the ferromagnetic model,
displaying mean field like behaviour.
\item[Fig. 4.1] The crossover in Binder's cumulant
for the overlap in the antiferromagnetic model
on $\phi^3$ graphs with $N=100,250,1000$.
\item[Fig. 4.2] The specific heat in the antiferromagnetic model,
which shows a characteristic cusp.
\item[Fig. 4.3] The linear susceptibility $\chi_M = dM / dH$
for the antiferromagnet.
\item[Fig. 4.4] The spin glass susceptibility $\chi_{sg}$
for the antiferromagnet.
\item[Fig. 4.5] A histogram of $P(q)$ at $\beta= -0.1$
which is deep in the paramagnetic phase. We have only plotted
the range $q=0,0.25$ and histograms for $n=250,1000$ for clarity.
\item[Fig. 4.6] A histogram of $P(q)$ at $\beta= -1.2$,
in the conjectured spin glass phase. We have plotted three graph sizes
to show the sharpening of the peak and the stability of the tail.
\item[Fig. 5.1] A histogram of $P(q)$ at $\beta=-1.9$
for the antiferromagnet on a $\phi^3$ planar (fat) graph
with 250 vertices. A histogram of $P(q)$ at $\beta= -1.3$
for a thin graph of similar size is also shown for comparison.
\end{description}
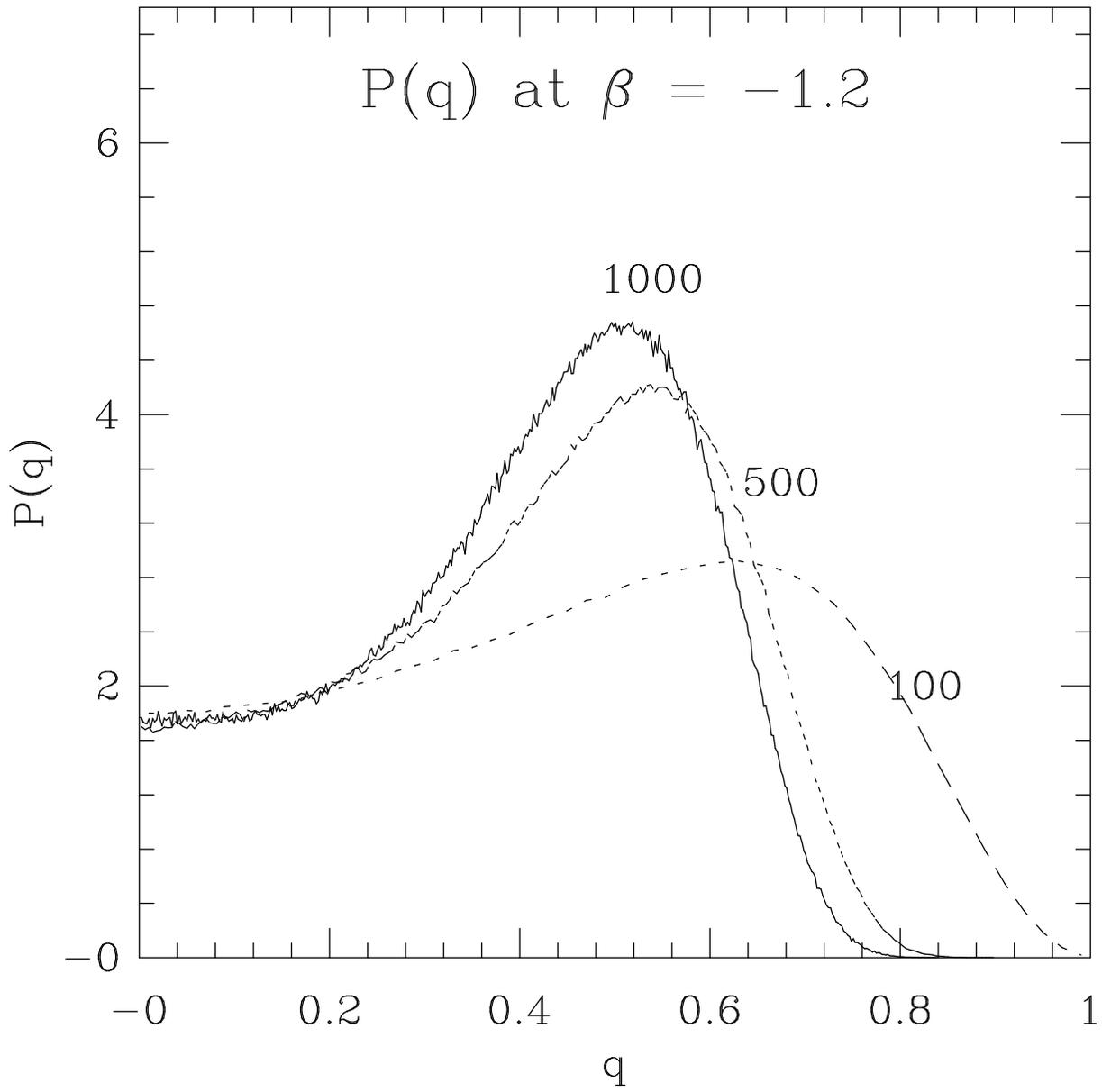
\end{document}